\documentclass[aps,prd,preprint,floats,superscriptaddress,showpacs]{revtex4}
\usepackage{graphicx}

\def \bea{\begin{eqnarray}}
\def \beq{\begin{equation}}
\def \bo{B^0}
\def \bra#1{\langle #1 |}

\def \eea{\end{eqnarray}}
\def \eeq{\end{equation}}

\def \ket#1{| #1 \rangle}

\def \ds{D_s}

\def \pr{\parallel}

\begin{document}

\preprint{ANL-HEP-PR-02-037, EFI-02-83, hep-ph/0206006}
\preprint{June 2002}

\title{$B \to D_s \pi$ and the tree amplitude in $B \to \pi^+ \pi^-$
  \footnote{To be submitted to Phys.\ Rev.\ D (Brief Reports).}}

\author{Cheng-Wei Chiang}
\email[e-mail: ]{chengwei@hep.uchicago.edu}
\affiliation{HEP Division, Argonne National Laboratory
9700 S. Cass Avenue, Argonne, IL 60439}
\affiliation{Enrico Fermi Institute and Department of Physics,
University of Chicago, 5640 S. Ellis Avenue, Chicago, IL 60637}
\author{Zumin Luo}
\email[e-mail: ]{zuminluo@midway.uchicago.edu}
\affiliation{Enrico Fermi Institute and Department of Physics, 
University of Chicago, 5640 S. Ellis Avenue, Chicago, IL 60637}
\author{Jonathan L.\ Rosner}
\email[e-mail: ]{rosner@hep.uchicago.edu}
\affiliation{Enrico Fermi Institute and Department of Physics, 
University of Chicago, 5640 S. Ellis Avenue, Chicago, IL 60637}

\date{\today}

\begin{abstract}
The recently-observed decay $B^0 \to D_s^+ \pi^-$ is expected to proceed
mainly by means of a tree amplitude in the factorization limit:  $B^0 \to \pi^-
{(W^+)}^*$, ${(W^+)}^* \to D_s^+$.  Under this assumption, we predict the
corresponding contribution of the tree amplitude to $B^0 \to \pi^+ \pi^-$.  We
indicate the needed improvements in data that will allow a useful estimate of
this amplitude with errors comparable to those accompanying other methods.
Since the factorization hypothesis for this process goes beyond that proved in
most approaches, we also discuss independent tests of this hypothesis.
\end{abstract}

\pacs{13.25.Hw, 14.40.Nd, 11.30.Hv.}

\maketitle



The two-body hadronic decay process $B^0 \to \pi^+ \pi^-$ has been of great
interest for a long time in the search for CP violation in $B$ decays.  Its
branching ratio, smaller than one typically estimates on the basis of
factorization and dominance of the tree-level amplitude $T$, may owe some
suppression to destructive interference between $T$ and the penguin amplitude
$P$ \cite{HHY,HSW,HY}.  This interference could provide information on both the
weak phase $\alpha = \phi_2$ and the relative strong phase of the tree and
penguin amplitudes.  Both quantities are helpful in testing the current picture
of CP violation based on phases in the Cabibbo-Kobayashi-Maskawa (CKM) matrix.
However, to answer the question of tree-penguin interference in $B^0 \to \pi^+
\pi^-$ requires improved knowledge of $|T|$ and $|P|$.  Since the tree
amplitude is the dominant contribution to $B^0 \to \pi^+ \pi^-$, better
knowledge of its magnitude is a key step toward such an improvement.
 
Within the factorization framework, if one simply takes form factor models and
computes the tree level amplitude of $B^0 \to \pi^+ \pi^-$, a significant error
will be obtained because of the large uncertainties in the form factor at large
recoil, $F_0(q^2 \to 0)$, and in $|V_{ub}|$.  Both of them have an error about
$\sim 25\%$, resulting in an error of more than $35\%$ on $|T|$.

In this article, we use the newly measured mode $\bo \to \ds^+ \pi^-$
\cite{Fabozzi:2002bv,Aubert} to estimate $\Gamma_{\rm tree}(\bo \to \pi^+
\pi^-)$. The uncertainty can be reduced because in the ratio of $\Gamma_{\rm
  tree}(\bo \to \pi^+ \pi^-) / \Gamma(\bo \to \ds^+ \pi^-)$ the dominant error
comes from the weak decay constant of $D_s$.  Within the next two years, the
CLEO-c program is expected to substantially improve the accuracy on various
charm sector parameter measurements, including $f_{D_s}$.  Therefore, we
propose an alternative method to determine $T$ for the $\bo \to \pi^+ \pi^-$
decay.  This method generally relies on a simple assumption about the pole
structure of the relevant $B \to \pi$ form factor to relate these two processes
at small and large recoil.  The same method can be applied to determining $T_P$
for $B^0 \to \rho^+ \pi^-$, where the subscript $P$ indicates that the
spectator quark goes into a pseudoscalar meson in the final state.


The $B \to \pi$ weak transition matrix element is conventionally parametrized
in the following way by two independent form factors:
\beq
\bra{\pi(p)} \bar{u}\gamma_{\mu}b \ket{B(p+q)} = 
\left(2p+q-q\frac{m_B^2-m_{\pi}^2}{q^2}\right)_{\mu}F_+(q^2) + 
q_{\mu}\frac{m_B^2-m_{\pi}^2}{q^2}F_0(q^2) .
\eeq
%

\begin{figure}
\includegraphics[height=4.5cm]{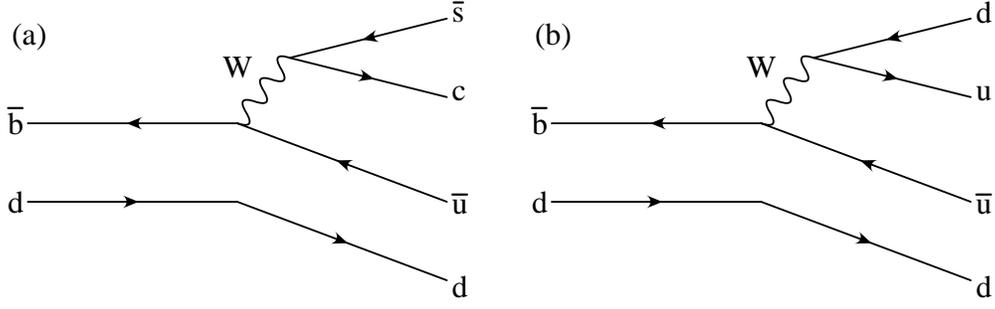}
\caption{Feynman diagrams for tree decays of a $\bo$ meson to $\ds^+ \pi^-$ and
  $\pi^+\pi^-$.
\label{fig:utrees}}
\end{figure}
Assuming factorization, the decay widths of $\bo \to \ds^+ \pi^-$ and $\bo \to
\pi^+ \pi^-$(tree) decay (as shown in Fig.~\ref{fig:utrees}) are given by:
\bea
\Gamma(\bo \to \ds^+ \pi^-) 
& = & \frac{G_F^2}{32\pi}|V_{ub}^* V_{cs}|^2f_{D_s}^2
      m_B\left(1-\frac{m_\pi^2}{m_B^2}\right)^2
\lambda(m_B^2,m_{D_s}^2,m_\pi^2) a_1^2 |F_0(m_{D_s}^2)|^2 ,
\label{dspi} \\
\Gamma_{\rm tree}(\bo \to \pi^+ \pi^-) 
& = & \frac{G_F^2}{32\pi}|V_{ub}^* V_{ud}|^2f_\pi^2
      m_B\left(1-\frac{m_\pi^2}{m_B^2}\right)^2
\lambda(m_B^2,m_\pi^2,m_\pi^2) a_1^2 |F_0(m_\pi^2)|^2 
\label{eqn:pipitree},
\eea
where $\lambda(a,b,c) \equiv \sqrt{a^2+b^2+c^2-2ab-2ac-2bc}$ and $a_1 \simeq 1$
is the Wilson coefficient.  Note that only $F_0(q^2)$ contributes in these two
decay modes.  To illustrate our method, we will use the form factor model
proposed in \cite{Becirevic:1999kt}, where $F_0(q^2)$ has the following single
pole structure:
\beq
F_0(q^2) = \frac{c_B(1-\alpha_B)}{1-q^2/(\beta_B m_{B^*}^2)} .
\eeq
A lattice calculation by Abada {\it et al} \cite{Abada:2000ty} gives
$F_0(0)=c_B(1-\alpha_B) = 0.26 \pm 0.05 \pm 0.04$ and $\beta_B=1.22 \pm
0.14^{+0.12}_{-0.03}$.

Let's define the ratio
\beq
\xi_B
\equiv \frac{{\cal B}_{\rm tree}(\bo \to \pi^+ \pi^-)}
              {{\cal B}(\bo \to \ds^+ \pi^-)} \nonumber
= \frac{\lambda(m_B^2,m_\pi^2,m_\pi^2)}{\lambda(m_B^2,m_{D_s}^2,m_\pi^2)}
  \left[ \frac{|V_{ud}|}{|V_{cs}|}
  \frac{f_\pi \, F_0(m_\pi^2)}{f_{D_s} \, F_0(m_{D_s}^2)} \right]^2 .
\eeq
In this ratio, the dependence upon $F_0(q^2 = 0)$ and $|V_{ub}|$ disappears
and, therefore, some large sources of uncertainty are avoided.  Neglecting the
errors on meson masses the pion decay constant, $f_\pi=131$ MeV, and the CKM
matrix elements (taking $|V_{ud}|=|V_{cs}|$ as suggested by unitarity), one
sees that the major error in $\xi_B$ comes from those of $f_{D_s}$ and
$\beta_B$.  In comparison with the error from $\beta_B$, which is given by the
lattice determination as mentioned earlier, a good portion of the uncertainty
in the ratio $\xi_B$ resides in the experimental determination of the $D_s$
decay constant.

Current experimental determination of $f_{D_s}$ uses the hadronic decay mode
$D_s^+ \to \phi \pi^+$ as a ``standard candle'' and measures the ratio of
${\cal B}(D_s^+ \to \ell^+ \nu_{\ell}) / {\cal B}(D_s^+ \to \phi \pi^+)$.
Therefore, the systematic error is dominated by the knowledge of ${\cal B}
(D_s^+ \to \phi \pi^+)$, which has a $25\%$ error \cite{PDG}.  Based on an
experimental average of rates for $D_s \to \mu \nu$ and $D_s \to \tau \nu$
\cite{Thaler:zc}, we will use $f_{D_s}=(270 \pm 16)\sqrt{{\cal B}(\ds^+ \to
  \phi \pi^+)/3.6\%}$ MeV for our numerical calculation, where the error is
purely statistical.  Here we single out the systematic error accompanying the
${\cal B}(\ds^+ \to \phi \pi^+)$ mode.  We will discuss its impact on the
precision determination of ${\cal B}_{\rm tree}(\bo \to \pi^+ \pi^-)$.

We first take the current value ${\cal B}(\ds^+ \to \phi \pi^+)=(3.6 \pm
0.9)\%$ \cite{PDG}.  Since $f_{D_s}^2$ is proportional to the ratio ${\cal
  B}(\ds^+ \to \mu^+(\tau^+) \nu)/{\cal B}(\ds^+ \to \phi \pi^+)$, we predict
\beq \label{eqn:ratio}
\xi_B
=(0.216 \pm 0.027)\left[\frac{3.6\%}{{\cal B}(\ds^+ \to \phi \pi^+)}\right] ,
\eeq
where we have combined the statistical error from $f_{D_s}$ and the error from
$\beta_B$, leaving the systematic error of $f_{D_s}$ in the square bracket.
Although $\beta_B$ has an error of $\sim 13\%$ by itself, it only results in a
$\sim 3\%$ error in $\xi_B$.  The statistical error of $f_{D_s}$, on the other
hand, gives a dominant $\sim 12\%$ error.

The BaBar collaboration \cite{Fabozzi:2002bv,Aubert} has recently measured the
product
\beq
{\cal B}(\bo \to \ds^+ \pi^-) \times {\cal B}(\ds^+ \to \phi \pi^+) 
= (1.11 \pm 0.37 \pm 0.22) \times 10^{-6}
\eeq
based on a data sample of $56.4~{\rm fb}^{-1}$ at $\Upsilon(4S)$ resonance,
where the first error is statistical and the second is systematic.
Using Eq.\ (\ref{eqn:ratio}), we immediately have
\beq
{\cal B}_{\rm tree}(\bo \to \pi^+ \pi^-)
= 6.7 (1 \pm 0.41) \times 10^{-6}
  \left[\frac{3.6\%}{{\cal B}(\ds^+ \to \phi \pi^+)}\right]^2~~~.
\eeq
Adding all the errors in quadrature, including that in ${\cal B}(\ds^+ \to \phi
\pi^+)$, we obtain
\beq
{\cal B}_{\rm tree}(\bo \to \pi^+ \pi^-) = 6.7(1 \pm 0.64)  \times 10^{-6},
\qquad
|T| = 2.6(1 \pm 0.32) \times 10^{-3}.
\eeq
This is in good agreement with the values obtained in \cite{Luo:2001ek} and
\cite{Rosner:2000ky}.  As stated before, direct calculation from
Eq.~(\ref{eqn:pipitree}) including the errors from $|V_{ub}|$ and $F_0(0)$ will
have an uncertainty in the branching ratio at least as big as $70\%$, which
would render the information useless.

As is obvious from the above analysis, the accuracy on the branching ratio of
$\ds^+ \to \phi \pi^+$ plays a crucial role in the determination of $|T|$.  It
is thus of great importance to lower its error.  The CLEO collaboration
proposes to explore the charm sector starting early 2003.  CLEO-c
\cite{Shipsey:2002kc} will be able to reach an accuracy of $1.9\%$ on ${\cal
  B}(\ds^+ \to \phi \pi^+)$ and in turn $1.7\%$ on $f_{D_s}$.  This will
improve our determination of $|T|$ considerably.  Moreover, if the data are
enlarged from the current 56.4 fb$^{-1}$ sample at BaBar to a combined BaBar
and Belle sample of 300 fb$^{-1}$, one expects to be able to bring down the
statistical error on ${\cal B}(\bo \to \ds^+ \pi^-) \times {\cal B}(\ds^+ \to
\phi \pi^+)$ by a factor of $\sim 2.3$.
With such reduced errors on $f_{D_s}$ and statistical error on the branching
ratio product, our knowledge of ${\cal B}_{\rm tree}(\bo \to \pi^+ \pi^-)$ can
be improved to give
\beq \label{eqn:T}
{\cal B}_{\rm tree}(\bo \to \pi^+ \pi^-) = 6.7(1 \pm 0.25)\times 10^{-6},
\qquad
|T| = 2.6(1 \pm 0.13) \times 10^{-3}.
\eeq
Now the error is dominated by the uncertainty in ${\cal B}(\bo \to \ds^+ \pi^-)
\times {\cal B}(\ds^+ \to \phi \pi^+)$.  Aside from reducing the statistical
error as mentioned before, it is also possible to reduce the systematic error
by, for example, improving the tagging techniques.

The anticipated error in Eq.\ (\ref{eqn:T}) is not as good as that (about 5\%)
foreseen in Ref.\ \cite{Luo:2001ek} on the basis of forthcoming studies of $B
\to \pi \ell \nu$.  Instead, it provides a cross-check of the factorization
hypothesis for the case in which the weak current produces a $D_s$.  Present
attempts to justify that hypothesis (see, e.g., \cite{BBNS}) do not expect it
to be valid when the weak current produces such a heavy color-singlet meson.
If we take the central values for the parameters appearing in Eq.\
(\ref{dspi}), however, we obtain ${\cal B}(\bo \to \ds^+ \pi^-) \simeq 2.9
\times 10^{-5}(|V_{ub}|/0.0036)^2$, consistent with the result presented in
Ref.\ \cite{Fabozzi:2002bv,Aubert}.  Therefore, current data do not indicate
any breakdown of factorization for $D_s$ or $D_s^*$ production by the weak
current, but more conclusive tests are needed \cite{LRfact}.


The above method can be similarly applied to the determination of the tree
amplitude $T_P$ in the $B^0 \to \rho^+ \pi^-$ decay using the experimental
branching ratio of $B^0 \to D_s^+ \pi^-$.  Using the same notation introduced
before,
\bea
\Gamma_{\rm tree}(\bo \to \rho^+ \pi^-)
= \frac{G_F^2}{32\pi} |V_{ub}^* V_{ud}|^2 f_{\rho}^2
  \left( \frac{\lambda(m_B^2,m_\rho^2,m_\pi^2)}{m_B} \right)^3
  a_1^2 |F_+(m_\rho^2)|^2 ,
\label{eqn:rhopitree}
\eea
where $f_{\rho} = 208$ MeV (see, for example, Ref.~\cite{Chiang:2001ir}).  We
consider an analogous ratio
\beq
\xi'_B
\equiv \frac{{\cal B}_{\rm tree}(\bo \to \rho^+ \pi^-)}
              {{\cal B}(\bo \to \ds^+ \pi^-)} \nonumber
= \frac{\lambda(m_B^2,m_\rho^2,m_\pi^2)^3}
       {m_B^4 \lambda(m_B^2,m_{D_s}^2,m_\pi^2)}
  \left( 1 - \frac{m_{\pi}^2}{m_B^2} \right)^{-2}
  \left[ \frac{|V_{ud}|}{|V_{cs}|}
  \frac{f_\rho \, F_+(m_\rho^2)}{f_{D_s} \, F_0(m_{D_s}^2)} \right]^2 .
\label{eqn:xiB2}
\eeq
This ratio generally involves additional model dependence because of
$F_+(q^2)$.  Ref.~\cite{Abada:2000ty} suggests the following parametrization:
\beq
F_+(q^2) = \frac{c_B (1 - \alpha_B)}
                {( 1 - q^2/m_{B^*}^2 )( 1 - \alpha_B q^2/m_{B^*}^2 )} ,
\eeq
where $\alpha_B$ has a value of $0.40 \pm 0.15 \pm 0.09$.  Again, $F_+(0)$
cancels with $F_0(0)$ in the ratio in Eq.~(\ref{eqn:xiB2}) and we find
\beq
\xi'_B = (0.541 \pm 0.018 \pm 0.004)\left[ 1 \pm \left( \frac{\Delta {\cal B}
(D_s \to \ell \nu)}{{\cal B}(D_s \to \ell \nu)} \right) \right]
         \left[\frac{3.6\%}{{\cal B}(\ds^+ \to \phi \pi^+)}\right] ,
\eeq
where the first error comes from $\beta_B$, the second error comes from
$\alpha_B$, and we have taken the central value of $f_{D_s}$ mentioned
previously.  Considering the same physics reach at CLEO-c and BaBar discussed
in the previous section, we obtain
\beq
{\cal B}_{\rm tree}(\bo \to \rho^+ \pi^-) = 16.7 (1 \pm 0.25)\times 10^{-6},
\qquad
|T_P| = 4.1 (1 \pm 0.13) \times 10^{-3},
\eeq
where the latter number agrees well with the estimate given in
Ref.~\cite{Chiang:2001ir}.  One may also contemplate estimating the above
quantities using the ratio ${\cal B}_{\rm tree}(\bo \to \rho^+ \pi^-) / {\cal
  B}(\bo \to D_s^{*+} \pi^-)$.  This has the advantage that the dependence on
$\beta_B$ disappears because both of the $VP$ modes involves only the form
factor $F_+$.  Although it is not observed yet, the branching ratio of $\bo \to
D_s^{*+} \pi^-$ is estimated to be of the same order as that of $\bo \to D_s^+
\pi^-$, except that it will have a bigger error due to the $\gamma$ detection
efficiency in $D_s^{*+}$ decay.  Currently, the BaBar group observes a
$2.2\sigma$ hint of the decay and sets an upper limit ${\cal B}(B^0 \to
D_s^{*+} \pi^-) < 4.3 \times 10^{-5}$ at 90\% confidence level \cite{Aubert}.
Nevertheless, a measurement of the former mode will still serve as a useful
check.

We have shown in this Article that within large experimental uncertainties the
present measurement of the branching ratio for $B^0 \to D_s^+ \pi^-$ is
compatible with the factorization hypothesis for production of the heavy meson
$D_s^+$ by the weak current.  Improvements in data [particularly in the
knowledge of ${\cal B}(D_s^+ \to \phi \pi^+)$] are pinpointed which will permit
a more conclusive test of this hypothesis.  It is also shown how observation of
the decay $B^0 \to D_s^{*+} \pi^-$ can provide a value of the tree amplitude in
$B^0 \to \rho^+ \pi^-$ which can be compared with that obtained through other
means (see, e.g., Ref.\ \cite{Chiang:2001ir}) to further test factorization in
this unexpected domain of its validity.


We thank Z. Ligeti for a discussion which stimulated the present analysis.
This work was supported in part by the U.\ S.\ Department of Energy, High
Energy Physics Division, under Grant No.\ DE-FG02-90ER-40560 and Contract
W-31109-ENG-38.

\def \ajp#1#2#3{Am.\ J. Phys.\ {\bf#1}, #2 (#3)}
\def \apny#1#2#3{Ann.\ Phys.\ (N.Y.) {\bf#1}, #2 (#3)}
\def \app#1#2#3{Acta Phys.\ Polonica {\bf#1}, #2 (#3)}
\def \arnps#1#2#3{Ann.\ Rev.\ Nucl.\ Part.\ Sci.\ {\bf#1}, #2 (#3)}
\def \art{and references therein}
\def \cmts#1#2#3{Comments on Nucl.\ Part.\ Phys.\ {\bf#1}, #2 (#3)}
\def \cn{Collaboration}
\def \cp89{{\it CP Violation,} edited by C. Jarlskog (World Scientific,
Singapore, 1989)}
\def \efi{Enrico Fermi Institute Report No.\ }
\def \epjc#1#2#3{Eur.\ Phys.\ J. C {\bf#1}, #2 (#3)}
\def \f79{{\it Proceedings of the 1979 International Symposium on Lepton and
Photon Interactions at High Energies,} Fermilab, August 23-29, 1979, ed. by
T. B. W. Kirk and H. D. I. Abarbanel (Fermi National Accelerator Laboratory,
Batavia, IL, 1979}
\def \hb87{{\it Proceeding of the 1987 International Symposium on Lepton and
Photon Interactions at High Energies,} Hamburg, 1987, ed. by W. Bartel
and R. R\"uckl (Nucl.\ Phys.\ B, Proc.\ Suppl., vol.\ 3) (North-Holland,
Amsterdam, 1988)}
\def \ib{{\it ibid.}~}
\def \ibj#1#2#3{~{\bf#1}, #2 (#3)}
\def \ichep72{{\it Proceedings of the XVI International Conference on High
Energy Physics}, Chicago and Batavia, Illinois, Sept. 6 -- 13, 1972,
edited by J. D. Jackson, A. Roberts, and R. Donaldson (Fermilab, Batavia,
IL, 1972)}
\def \ijmpa#1#2#3{Int.\ J.\ Mod.\ Phys.\ A {\bf#1}, #2 (#3)}
\def \ite{{\it et al.}}
\def \jhep#1#2#3{JHEP {\bf#1}, #2 (#3)}
\def \jpb#1#2#3{J.\ Phys.\ B {\bf#1}, #2 (#3)}
\def \lg{{\it Proceedings of the XIXth International Symposium on
Lepton and Photon Interactions,} Stanford, California, August 9--14 1999,
edited by J. Jaros and M. Peskin (World Scientific, Singapore, 2000)}
\def \lkl87{{\it Selected Topics in Electroweak Interactions} (Proceedings of
the Second Lake Louise Institute on New Frontiers in Particle Physics, 15 --
21 February, 1987), edited by J. M. Cameron \ite~(World Scientific, Singapore,
1987)}
\def \kdvs#1#2#3{{Kong.\ Danske Vid.\ Selsk., Matt-fys.\ Medd.} {\bf #1},
No.\ #2 (#3)}
\def \ky85{{\it Proceedings of the International Symposium on Lepton and
Photon Interactions at High Energy,} Kyoto, Aug.~19-24, 1985, edited by M.
Konuma and K. Takahashi (Kyoto Univ., Kyoto, 1985)}
\def \mpla#1#2#3{Mod.\ Phys.\ Lett.\ A {\bf#1}, #2 (#3)}
\def \nat#1#2#3{Nature {\bf#1}, #2 (#3)}
\def \nc#1#2#3{Nuovo Cim.\ {\bf#1}, #2 (#3)}
\def \nima#1#2#3{Nucl.\ Instr.\ Meth. A {\bf#1}, #2 (#3)}
\def \np#1#2#3{Nucl.\ Phys.\ {\bf#1}, #2 (#3)}
\def \npbps#1#2#3{Nucl.\ Phys.\ B Proc.\ Suppl.\ {\bf#1}, #2 (#3)}
\def \npps#1#2#3{Nucl.\ Phys.\ Proc.\ Suppl.\ {\bf#1}, #2 (#3)}
\def \os{XXX International Conference on High Energy Physics, Osaka, Japan,
July 27 -- August 2, 2000}
\def \PDG{Particle Data Group, D. E. Groom \ite, \epjc{15}{1}{2000}}
\def \pisma#1#2#3#4{Pis'ma Zh.\ Eksp.\ Teor.\ Fiz.\ {\bf#1}, #2 (#3) [JETP
Lett.\ {\bf#1}, #4 (#3)]}
\def \pl#1#2#3{Phys.\ Lett.\ {\bf#1}, #2 (#3)}
\def \pla#1#2#3{Phys.\ Lett.\ A {\bf#1}, #2 (#3)}
\def \plb#1#2#3{Phys.\ Lett.\ B {\bf#1}, #2 (#3)}
\def \pr#1#2#3{Phys.\ Rev.\ {\bf#1}, #2 (#3)}
\def \prc#1#2#3{Phys.\ Rev.\ C {\bf#1}, #2 (#3)}
\def \prd#1#2#3{Phys.\ Rev.\ D {\bf#1}, #2 (#3)}
\def \prl#1#2#3{Phys.\ Rev.\ Lett.\ {\bf#1}, #2 (#3)}
\def \prp#1#2#3{Phys.\ Rep.\ {\bf#1}, #2 (#3)}
\def \ptp#1#2#3{Prog.\ Theor.\ Phys.\ {\bf#1}, #2 (#3)}
\def \rmp#1#2#3{Rev.\ Mod.\ Phys.\ {\bf#1}, #2 (#3)}
\def \rp#1{~~~~~\ldots\ldots{\rm rp~}{#1}~~~~~}
\def \si90{25th International Conference on High Energy Physics, Singapore,
Aug. 2-8, 1990}
\def \slc87{{\it Proceedings of the Salt Lake City Meeting} (Division of
Particles and Fields, American Physical Society, Salt Lake City, Utah, 1987),
ed. by C. DeTar and J. S. Ball (World Scientific, Singapore, 1987)}
\def \slac89{{\it Proceedings of the XIVth International Symposium on
Lepton and Photon Interactions,} Stanford, California, 1989, edited by M.
Riordan (World Scientific, Singapore, 1990)}
\def \smass82{{\it Proceedings of the 1982 DPF Summer Study on Elementary
Particle Physics and Future Facilities}, Snowmass, Colorado, edited by R.
Donaldson, R. Gustafson, and F. Paige (World Scientific, Singapore, 1982)}
\def \smass90{{\it Research Directions for the Decade} (Proceedings of the
1990 Summer Study on High Energy Physics, June 25--July 13, Snowmass, Colorado),
edited by E. L. Berger (World Scientific, Singapore, 1992)}
\def \tasi{{\it Testing the Standard Model} (Proceedings of the 1990
Theoretical Advanced Study Institute in Elementary Particle Physics, Boulder,
Colorado, 3--27 June, 1990), edited by M. Cveti\v{c} and P. Langacker
(World Scientific, Singapore, 1991)}
\def \yaf#1#2#3#4{Yad.\ Fiz.\ {\bf#1}, #2 (#3) [Sov.\ J.\ Nucl.\ Phys.\
{\bf #1}, #4 (#3)]}
\def \zhetf#1#2#3#4#5#6{Zh.\ Eksp.\ Teor.\ Fiz.\ {\bf #1}, #2 (#3) [Sov.\
Phys.\ - JETP {\bf #4}, #5 (#6)]}
\def \zpc#1#2#3{Zeit.\ Phys.\ C {\bf#1}, #2 (#3)}
\def \zpd#1#2#3{Zeit.\ Phys.\ D {\bf#1}, #2 (#3)}


\end{document}